\documentclass[aps,prd,onecolumn,groupedaddress,nofootinbib,amssymb]{revtex4}

\usepackage[dvips]{graphicx}
\usepackage{amssymb}
\usepackage{amsmath}
\usepackage{graphicx,,color}
\usepackage{amsfonts}
\usepackage{bm}

\newcommand\e{\mathrm{e}}
\newcommand\nn{\nonumber \\}
\newcommand{\be}{\begin{equation}}
\newcommand{\ee}{\end{equation}}

\allowdisplaybreaks[4]

\begin{document}

\tolerance=5000

\title{How fundamental is entropy? \\
From non-extensive statistics and 
black hole physics to the holographic dark universe}

\author{Shin'ichi~Nojiri$^{1,2}$, Sergei~D.~Odintsov$^{3,4,5}$, Valerio 
Faraoni$^{6}$} 
\affiliation{$^1$Department of Physics, Nagoya University, 
Nagoya 464-8602, Japan\\ 
$^2$Kobayashi-Maskawa Institute for 
the Origin of Particles and the Universe, Nagoya University, Nagoya 
464-8602, Japan\\ 
$^3$Instituci\'{o} Catalana de Recerca i Estudis 
Avan\c{c}ats (ICREA), Passeig Llu\'{i}s Companys, 23, 08010 Barcelona, 
Spain\\ 
$^4$ Institute of Space Sciences (IEEC-CSIC), C. Can Magrans s/n , 08093 
Barcelona, Spain \\
$^5$ Int.~Lab.~Theor.~Cosmology, Tomsk State University of Control 
Systems and Radioelectronics(TUSUR), 634050 
Tomsk, Russia\\
$^6$ Department of Physics \& Astronomy, 
Bishop's University, 2600 College Street, Sherbrooke, Qu\'ebec, Canada J1M 
1Z7
 }

\begin{abstract}

We propose a new entropy construct that generalizes the Tsallis, R\'enyi, 
Sharma-Mittal, Barrow, Kaniadakis, and Loop Quantum Gravity entropies and 
reduces to the Bekenstein-Hawking entropy in a certain limit. This 
proposal is applied to the Schwarzschild black hole and to spatially 
homogeneous and isotropic cosmology, where it is shown that it can 
potentially describe inflation and/or holographic dark energy.

\end{abstract} 
\maketitle

\section{Introduction}
\label{sec:1}

Since the golden days of classical thermodynamics, entropy has been viewed 
as a unique, universal, and fundamental quantity playing one of the most 
important roles in physics. However, with the development of quantum 
physics, quantum field theory, quantum gravity, and non-extensive 
thermodynamics, it progressively became clear that entropy is not unique 
and is not as fundamental. Somehow, it depends on the physical system 
under consideration and it changes across physical theories. Eventually, 
this property implies that we do not fundamentally understand what 
physical entropy is and that the basic principles underlying its 
construction should be revisited from a critical point of view.

Over the years, we have witnessed the appearance of a variety of entropies 
in many classical and quantum systems. Some of these entropy concepts were 
introduced starting from very different points of view, each one of which 
is certainly valid on its own, originating a legitimate proposal in the 
context in which it was conceived. These entropies highlight different 
aspects of natural phenomena or different approaches to physical theories. 
This proliferation of entropy notions makes it clear that entropy may not 
be a uniquely defined concept and that there may exist even more 
entropies than those already proposed.

One of the main surprises of theoretical physics in the 1970s was that 
black holes are not cold objects but have entropy and temperature. 
Bekenstein's association of entropy with black holes, proportional to 
the black hole horizon area \cite{Bekenstein:1973ur}, remained odd and 
inconclusive until Hawking discovered that the Schwarzschild black hole 
(and, by extension, all black holes) radiate quanta of quantum fields 
living on that spacetime, emitting a blackbody spectrum at a temperature 
$T_\mathrm{H}=\frac{1}{8\pi G M}$, where $M$ is the black hole mass 
\cite{Hawking:1975vcx}. The discovery of the Hawking temperature made 
sense of Bekenstein's black hole entropy and paved the way for the 
development of black hole thermodynamics (\cite{Bardeen:1973gs}, see 
\cite{Wald:1999vt, Carlip:2014pma} for reviews).

One puzzling feature of the Bekenstein-Hawking entropy was, from its  
beginnings, that it is not proportional to the black hole volume, as 
familiar in classical thermodynamics, but rather it is proportional to the 
black hole horizon area. In classical thermodynamics, the entropy of a system 
is proportional to its mass and its volume and is an extensive and 
additive quantity; the fundamental reason why black hole entropy is 
instead non-extensive remains shrouded in mystery \cite{Tsallis:2012js}. 
Given the elusive nature of the origin of this entropy, it is not 
surprising that recent literature \cite{Bialas:2008fa, Tsallis:2012js, 
Huang:2014pda, Brustein:2014iha, Nishioka:2014mwa, Czinner:2015eyk, 
Dong:2016fnf, Wen:2016itq, Czinner:2017tjq, Qolibikloo:2018wqw, 
Johnson:2018bma, Tannukij:2020njz, Promsiri:2020jga, Samart:2020klx, 
Ren:2020djc, Mejrhit:2020dpo, Nakarachinda:2021jxd, Abreu:2021avp} 
contemplates alternatives, replacing the Bekenstein-Hawking entropy with 
other constructs based on non-extensive statistics, including the R\'enyi 
\cite{renyi} and Tsallis \cite{tsallis} non-extensive entropies (a better 
terminology would ``non-additive'' entropies). Other notable notions of 
entropy which have been studied recently are the Barrow entropy arising 
from the attempt to take into account the quantum spacetime foam 
\cite{Barrow:2020tzx}, and the Sharma-Mittal 
\cite{SayahianJahromi:2018irq} and Kaniadakis 
\cite{Kaniadakis:2005zk,Drepanou:2021jiv} proposals. Since entropy, 
temperature, internal energy, and heat transferred are related by the 
first law of thermodynamics, changing the notion of entropy entails 
changes in these other quantities, usually jeopardizing the first law, as 
discussed in \cite{Nojiri:2021czz}.

What is more, horizons are not a prerogative of black holes but appear 
also in cosmology, hence horizon thermodynamics was extended to 
cosmological 
horizons. Incidentally, realistic black holes live in the universe and are 
not asymptotically flat: the simple addition of a cosmological constant to 
the Einstein equations change their black hole solutions and their 
thermodynamics becomes richer. However, cosmology itself poses several 
interesting questions, one of the most important being how to explain the 
present acceleration of the cosmic expansion discovered in 1998 with Type 
Ia supernovae. For 
this purpose, many scenarios of dark energy and modified gravity have 
been proposed and are being tested and/or constrained as newer 
cosmological observations become available.  Among the many scenarios 
advanced in the cosmology literature, the holographic dark energy proposal 
\cite{Li:2004rb,Li:2011sd,Wang:2016och, 
Nojiri:2005pu,Enqvist:2004xv,Zhang:2005yz,Guberina:2005fb,Elizalde:2005ju, 
Ito:2004qi,Gong:2004cb,Saridakis:2007cy,Gong:2009dc,BouhmadiLopez:2011xi,Malekjani:2012bw, 
Khurshudyan:2014axa,Khurshudyan:2016gmb,Landim:2015hqa, 
Gao:2007ep, 
Li:2008zq,Anagnostopoulos:2020ctz,Zhang:2005hs,Li:2009bn,Feng:2007wn,Lu:2009iv, 
Micheletti:2009jy,Huang:2004wt,Mukherjee:2017oom,Nojiri:2017opc, 
Sharif:2019seo,Jawad:2018juh,Bolanos:2021wmn,Granda:2018biv,Chakraborty:2020jsq,Paul:2019hys})  
is directly related to entropy. Therefore, replacing the notion of entropy 
used in physics has a direct impact on this scenario.

Beginning from the realization that the various entropies alternative to 
the Bekenstein-Hawking one proposed in the literature share certain 
properties, including the fact that they reduce to the Bekenstein-Hawking 
entropy in a certain limit, we investigate two new generalized entropies 
that contain all these previous proposals as special cases. The first of 
these new entropies features six parameters, but we provide also a 
simplified version containing only three parameters, which we then apply 
to black holes and to holographic dark energy in cosmology. Already the 
simplified proposal has the potential of generating two vastly different 
energy scales associated with inflation or with the present acceleration.

In the next section, we review various entropies studied in the recent 
literature and introduce the generalized entropy that reproduces them for 
special parameter values. A simplified version of this generalized entropy 
is then applied to black holes in Sec.~\ref{sec:3} and to holographic dark 
energy in Sec.~\ref{sec:4}. We mostly follow the notation of 
Ref.~\cite{Wald}, using units in which the speed of light $c$, the 
Boltzmann constant $K_B$, and the reduced Planck constant $\hbar$ are 
unity, $G$ is Newton's constant, $\kappa^2 \equiv 8\pi G$, while the 
metric signature is $({-}{+}{+}{+})$.

\section{Possible generalizations of known entropies}
\label{sec:2}

Let us begin with the standard thermodynamical entropy of black hole 
physics, one of the most far-reaching applications of entropy that led to 
the development of black hole thermodynamics 
\cite{Bardeen:1973gs,Wald:1999vt,Carlip:2014pma}. The Bekenstein-Hawking 
entropy is \cite{Bekenstein:1973ur,Hawking:1975vcx}
\begin{align}
\label{SAdS10}
\mathcal{S} = \frac{A}{4G}\,,
\end{align}
where $A\equiv 4\pi {r_\mathrm{h}}^2$ is the area of the horizon and 
$r_\mathrm{h}$ is the horizon radius (using the areal radius as the radial 
coordinate). This proposal, however, is not unique.  Indeed, depending on 
the system under consideration, different entropies may be introduced. Let 
us recall some of the entropy concepts proposed thus far.

\begin{itemize}

\item The Tsallis entropy \cite{tsallis} (see also 
\cite{Ren:2020djc,Nojiri:2019skr}) appears in the study of non-extensive 
statistics for systems with long range interactions, in which the 
partition function diverges and the standard Boltzmann-Gibbs entropy 
becomes inadequate; it is  
\begin{align}
\label{TS1}
\mathcal{S}_\mathrm{T} = \frac{A_0}{4G} \left( \frac{A}{A_0} 
\right)^\delta \,,
\end{align}
where $A_0$ is a constant with the dimensions of an area and $\delta$ is 
a dimensionless parameter that quantifies the non-extensivity. The 
standard Bekenstein-Hawking entropy~(\ref{SAdS10}) is recovered for $\delta = 1$. 

\item The R{\'e}nyi entropy (\cite{renyi}, see also 
\cite{Czinner:2015eyk,Tannukij:2020njz,Promsiri:2020jga,Samart:2020klx}) 
is defined as 
\begin{align}
\label{RS1}
\mathcal{S}_\mathrm{R}=\frac{1}{\alpha} \ln \left( 1 + \alpha 
\mathcal{S} \right) 
\end{align}
where $\mathcal{S}$ is identified with the Bekenstein-Hawking 
entropy~(\ref{SAdS10}), and contains a parameter $\alpha$.  The R\'enyi 
entropy was proposed as an index specifying the amount of information and, 
originally, had no relation with the statistics of physical systems.

\item The Sharma-Mittal entropy is \cite{SayahianJahromi:2018irq}
\begin{align}
\label{SM}
\mathcal{S}_\mathrm{SM} = \frac{1}{R}\left[ \left(1 + \delta 
\, \mathcal{S}_\mathrm{T} \right)^{R/\delta} - 1 \right]
\end{align}
where $\mathcal{S}_\mathrm{T}$ is the Tsallis entropy, while $R$ and 
$\delta$ are free phenomenological parameters to be determined by 
the best-fitting of  experimental data. The Sharma-Mittal entropy can be 
seen as a combination of the R{\'e}nyi and Tsallis entropies.

\item The Barrow entropy is \cite{Barrow:2020tzx}
\begin{align}
\label{barrow}
\mathcal{S}_\mathrm{B} = \left( 
\frac{A}{A_\mathrm{Pl}}\right)^{1+ \, \Delta/2}\,;
\end{align}
here $A$ is the usual black hole horizon area and $A_\mathrm{Pl}\equiv 4G$ 
is the Planck area. Formally, the Barrow entropy resembles the Tsallis 
non-extensive entropy but the physical principles underlying its 
introduction are radically different. The Barrow entropy was proposed as a 
toy model for the possible effects of quantum gravitational spacetime 
foam.  The quantum-gravitational deformation is quantified by the new 
exponent $\Delta$. The Barrow entropy reduces to the standard 
Bekenstein-Hawking entropy in the limit 
$\Delta \rightarrow 0$, while $\Delta = 1$ corresponds to maximal 
deformation.

\item The Kaniadakis entropy \cite{Kaniadakis:2005zk,Drepanou:2021jiv} 
\begin{align}
\label{kani}
\mathcal{S}_\mathrm{K} = \frac{1}{K} \sinh \left(K \mathcal{S} \right)\, 
,
\end{align}
reproduces the Bekenstein-Hawking entropy in the limit $K \to 0$ of its 
parameter $K$. It can be regarded as a generalization of the 
Boltzmann-Gibbs entropy arising in relativistic statistical systems 
\cite{Kaniadakis:2005zk,Drepanou:2021jiv}.

\item Non-extensive statistical mechanics in Loop Quantum Gravity gives 
the entropy \cite{Majhi:2017zao,Czinner:2015eyk,Mejrhit:2019oyi, Liu:2021dvj}
\be
\mathcal{S}_q= \frac{1}{1-q} \left[ \mbox{e}^{ (1-q)\Lambda(\gamma_0) {\cal 
S} } -1 \right] \,,\label{LQGentropy}
\ee
where the entropic index $q$ quantifies how the probability of frequent 
events is enhanced relatively to infrequent ones, 
\be
\Lambda( \gamma_0) = \frac{ \ln 2}{\sqrt{3} \, \pi \gamma_0} \,,
\ee
and $\gamma_0$ is the Barbero-Immirzi parameter, which is usually assumed 
to take one of the two values $ \frac{\ln2}{\pi \sqrt{3}}$ or $\frac{\ln 
3}{2\pi \sqrt{2} }$, depending on the gauge group used in Loop Quantum 
Gravity. However, $\gamma_0 $ is a free parameter in scale-invariant 
gravity \cite{Veraguth:2017uwp, Wang:2018bdg, Wang:2019ryx}. With the 
first choice of $\gamma_0$, $\Lambda(\gamma_0)$ becomes unity and the 
entropy~(\ref{LQGentropy}) reduces to the Bekenstein-Hawking one in the 
limit $q\rightarrow 1$, which corresponds to extensive statistical 
mechanics. This Loop Quantum Gravity entropy~(\ref{LQGentropy}) was 
applied to black holes in \cite{Majhi:2017zao, Czinner:2015eyk, 
Mejrhit:2019oyi} and to cosmology in \cite{Liu:2021dvj}. 
\end{itemize}

The above entropies share the following properties: 

\begin{enumerate}

\item {\it Generalized third law:} All these entropies vanish when the 
Bekenstein-Hawking entropy vanishes. In the third law of standard 
thermodynamics for closed systems in thermodynamic equilibrium, the 
quantity $\e^\mathcal{S}$ expresses the number of  
states, or the volume of these states, and therefore the entropy 
$\mathcal{S}$ vanishes when the temperature does because the ground 
(vacuum) state should be unique. By contrast, the Bekenstein-Hawking 
entropy $\mathcal{S}$ diverges when the temperature $T$ vanishes and it goes 
to zero at infinite temperature. However, requiring any generalized 
entropy to vanish when the Bekenstein-Hawking entropy $\mathcal{S}$ vanishes 
could be a natural requirement.

\item {\it Monotonically increasing functions:} All the above entropies 
are monotonically increasing functions of the Bekenstein-Hawking entropy 
$\mathcal{S}$.

\item {\it Positivity:} All the above entropies are positive, as is the 
Bekenstein-Hawking entropy~(\ref{SAdS10}). This is natural because 
$\e^\mathcal{S}$ corresponds to the number of states (or to the volume of 
these states), which is greater than unity.

\item {\it Bekenstein-Hawking limit:} All the above entropies reduce to 
the Bekenstein-Hawking entropy~(\ref{SAdS10}) in an appropriate limit.

\end{enumerate}

In the preceding expressions, all entropies are functions of the 
Bekenstein-Hawking entropy~(\ref{SAdS10}). In this sense, the most general 
entropy $\mathcal{S}_\mathrm{G}$ would be a function of the 
Bekenstein-Hawking entropy $\mathcal{S}$,
\begin{align}
\label{general}
\mathcal{S}_\mathrm{G}=\mathcal{S}_\mathrm{G} \left( \mathcal{S} \right) 
\,,
\end{align}
subject to certain natural requirements: we require the general entropy 
$\mathcal{S}_\mathrm{G}$ to possess  
the above properties. 

An example of such an entropy construct containing  
six parameters $\left( \alpha_\pm , \beta_\pm , \gamma_\pm \right)$ 
could be 
\begin{align}
\label{general1}
\mathcal{S}_\mathrm{G} \left( \alpha_\pm, \beta_\pm, \gamma_\pm \right)
=  \frac{1}{\alpha_+ + \alpha_-}
\left[ \left( 1 + \frac{\alpha_+}{\beta_+} \, \mathcal{S}^{\gamma_+} 
\right)^{\beta_+} - \left( 1 + \frac{\alpha_-}{\beta_-} 
\, \mathcal{S}^{\gamma_-} \right)^{-\beta_-} \right] \,,
\end{align}
where we assume all the parameters $\left( \alpha_{\pm} , \beta_{\pm},  
\gamma_{\pm} \right) $ to be positive. First, we show that the entropy 
$\mathcal{S}_\mathrm{G} \left( \alpha_\pm, \beta_\pm, \gamma_\pm \right)$ 
reduces to the entropies~(\ref{TS1}), (\ref{RS1}), (\ref{SM}), 
(\ref{barrow}), (\ref{kani}), and (\ref{LQGentropy}) already presented for 
appropriate choices of the parameter values.

\begin{itemize}
 
\item In the limit $\alpha_+=\alpha_- \to 0$, the choice  
$\gamma_{+}= \gamma_{-} \equiv \gamma$ gives 
\begin{align}
\label{general2}
\mathcal{S}_\mathrm{G} \left( \alpha_\pm\to 0, \beta_\pm, \gamma \right)
\to \mathcal{S}^\gamma \, .
\end{align}
If we further choose $\gamma=\delta$ or $\gamma= 1 + \Delta/2 $, the 
Tsallis entropy~(\ref{TS1}) or the Barrow entropy (\ref{barrow}) are 
reproduced, respectively.

\item The parameter  choice $\alpha_-=0$ yields
\begin{align}
\label{general3}
\mathcal{S}_\mathrm{G} \left( \alpha_+, \alpha_-=0, \beta_\pm, 
\gamma_+=1, \gamma_- \right)
= \frac{1}{\alpha_+}\left[ \left( 1 + 
\frac{\alpha_+}{\beta_+}\mathcal{S}^{\gamma_+} \right)^{\beta_+} - 1 
\right] \,.
\end{align}

Then, writing $\alpha_+=R$, $\beta_+ = R/\delta$, and 
$\gamma_+=\delta$, one obtains the Sharma-Mittal entropy~(\ref{SM}).

\item In Eq.~(\ref{general3}), if we further take the limit  
$\alpha_+ \rightarrow 0 $ simultaneously with $\beta_+ \rightarrow 0 $  
keeping $\alpha \equiv \alpha_+ / \beta_+ $ finite, and we choose 
$\gamma_+=1$, we obtain 
\begin{align}
\label{general4}
\mathcal{S}_\mathrm{G} &\, \left( \alpha_+\to 0, \alpha_-=0, \beta_+ \to 
0, \beta_-, \gamma_+=1, \gamma_-;  \alpha \equiv 
\frac{\alpha_+}{\beta_+} \, \mbox{finite} \right) \nn
& \to  \frac{1}{\alpha_+}\left[ \e^{\beta_+\ln \left( 1 + 
\frac{\alpha_+}{\beta_+}\mathcal{S} \right)}- 1 \right] 
\simeq \frac{1}{\alpha_+}\left[ 1 + \beta_+\ln \left( 1 + 
\frac{\alpha_+}{\beta_+}\mathcal{S} \right) - 1 \right] 
= \frac{\beta_+}{\alpha_+}\ln \left( 1 + 
\frac{\alpha_+}{\beta_+}\mathcal{S} \right) \nn
& \equiv \frac{1}{\alpha}\ln \left( 1 + \alpha \mathcal{S} \right) \, ,
\end{align}
which reproduces the R{\'e}nyi entropy~(\ref{RS1}). 

\item Taking the limit $\beta_\pm\to 0$, choosing  
$\gamma_\pm=1$,  and writing $\alpha_\pm = K$, the general 
entropy~(\ref{general1}) reduces to the Kaniadakis one~(\ref{kani}), 
\begin{align}
\label{general5}
\mathcal{S}_\mathrm{G} \left( \alpha_\pm=K, \beta_\pm\to 0, \gamma_\pm=1 
\right) \to  \frac{1}{2K}\left( \e^{K\mathcal{S}} - \e^{-K\mathcal{S}} 
\right)  = \frac{1}{K} \sinh \left(K \mathcal{S} \right) \, .
\end{align}

\item Finally, taking $\alpha_{-}=0 $ and $\gamma_{+}=1$ in the 
generalized entropy~(\ref{general1}), one obtains
\be
\mathcal{S}_G =\frac{1}{\alpha_{+}} \left[ \mbox{e}^{ \beta_{+} \ln \left( 
1+\frac{\alpha_{+}}{\beta_{+}} \, \mathcal{S} \right)} -1 \right] 
\ee
and the further limit $\beta_{+} \rightarrow +\infty $ in conjunction 
with $\alpha=1-q \, $ yields 
\be
\mathcal{S}_G \approx \frac{1}{1-q} \left[ \mbox{e}^{(1-q) \mathcal{S}} -1 
\right] 
\ee
corresponding to $\Lambda( \gamma_0 ) =1$ in the Loop Quantum Gravity 
entropy~(\ref{LQGentropy}), and which reduces to  the 
Bekenstein-Hawking entropy $\mathcal{S}$ as $q\to 1$.

\end{itemize}

It is straightforward to check that the entropy~$\mathcal{S}_\mathrm{G} 
\left( \alpha_\pm, \beta_\pm, \gamma_\pm \right)$ in Eq.~(\ref{general1}) 
satisfies  the generalized third law, that is, $\mathcal{S}_\mathrm{G} 
\left( 
\alpha_\pm, \beta_\pm, \gamma_\pm \right)\to 0$ when $\mathcal{S} \to 0$. 
The entropy $\mathcal{S}_\mathrm{G} \left( \alpha_\pm, \beta_\pm, 
\gamma_\pm \right)$ is a monotonically increasing function of 
$\mathcal{S}$ 
because both $\left( 1 + \, \frac{\alpha_+}{\beta_+} 
\, \mathcal{S}^{\gamma_+} 
\right)^{\beta_+}$ and $- \left( 1 + \,  
\frac{\alpha_-}{\beta_-} \, \mathcal{S}^{\gamma_-} \right)^{-\beta_-}$ are 
monotonically increasing functions of $\mathcal{S}$, given that all the 
parameters $\left( \alpha_\pm , \beta_\pm , \gamma_\pm \right)$ are 
assumed to be 
positive, and their sum is also monotonically increasing. Positivity is 
satisfied because $\mathcal{S}_\mathrm{G} 
\left( \alpha_\pm, \beta_\pm, \gamma_\pm \right)= 0$ when $\mathcal{S} = 
0$ and $\mathcal{S}_\mathrm{G} \left( \alpha_\pm, \beta_\pm, \gamma_\pm 
\right)$ is a strictly increasing function of $\mathcal{S}$. 

It is clear that there exists a limit of $\mathcal{S}_\mathrm{G} \left( 
\alpha_\pm, \beta_\pm, \gamma_\pm \right)$ to the Bekenstein-Hawking 
entropy because $\mathcal{S}_\mathrm{G} $ reduces to the 
entropies~(\ref{TS1}), (\ref{RS1}), (\ref{SM}), (\ref{barrow}), 
(\ref{kani}), and~(\ref{LQGentropy}), which have the required limiting 
behaviour. More explicitly, we have 
\begin{equation} 
\lim_{ \alpha_\pm\to 0} \, \mathcal{S}_\mathrm{G} \left( \alpha_\pm , 
\beta_\pm, \gamma_\pm \right)=\mathcal{S} \,. 
\end{equation}

We may also consider the three-parameter entropy-like quantity 
\begin{align}
\label{general6}
\mathcal{S}_\mathrm{G} \left( \alpha, \beta, \gamma \right)
= \frac{1}{\gamma} \left[ \left( 1 + \frac{\alpha}{\beta} \, \mathcal{S} 
\right)^\beta - 1 \right] \,,
\end{align}
where we assume again the parameters $ \left( \alpha , \beta, \gamma 
\right)$ to be  positive. 
When $\gamma$ and $\alpha$ coincide, the expression~(\ref{general6}) 
reduces to the Sharma-Mittal entropy~(\ref{SM}) with 
$\mathcal{S}_\mathrm{T}=\mathcal{S}$, that is, $\delta=1$. By writing  
$\gamma = \left( \alpha/\beta \right)^{\beta} $, the 
limit $\alpha\to \infty$ yields 
\begin{align}
\label{general7}
\lim_{\alpha\to \infty} \mathcal{S}_\mathrm{G} \left( \alpha, \beta, 
\gamma=\left( \frac{\alpha}{\beta} \right)^\beta \right)
= \mathcal{S}^\beta \, .
\end{align}
The choices $\beta=\delta$ and $\beta= 1 + \Delta/2$ give the 
Tsallis entropy~(\ref{TS1}) and the Barrow entropy~(\ref{barrow}), 
respectively. 
If, instead, we consider the limit in which $\alpha\rightarrow 0 $ and 
$\beta \rightarrow 0$ simultaneously while keeping $ \alpha/\beta $  
finite, as in Eq.~(\ref{general4}), we 
obtain the R{\'e}nyi entropy (\ref{RS1}) by replacing $\alpha/\beta$ 
with $\alpha$ and choosing $\gamma=\alpha$,  
\begin{align}
\label{general8}
\mathcal{S}_\mathrm{G} \left( \alpha\to 0, \beta\to 0, \gamma; 
\frac{\alpha}{\beta} \, \mbox{finite}  \right)
\to \frac{1}{\gamma} \ln \left( 1 + \frac{\alpha}{\beta} \, \mathcal{S} 
\right)  = \frac{1}{\alpha}\ln \left( 1 + \alpha \mathcal{S} \right) \,.
\end{align}

Another possibility consists of taking the limit $\beta\to \infty$ in 
conjunction with $\gamma=\alpha$, which leads to the new 
type of expression 
\begin{align}
\label{general9}
\mathcal{S}_\mathrm{G} \left( \alpha, \beta\to \infty, \gamma \right)
\to \frac{1}{\gamma} \left( \e^{\alpha\mathcal{S}} - 1 \right) \, .
\end{align}
It is again straightforward to check that~(\ref{general6}) 
satisfies all the conditions characterizing the generalized third law: 
monotonically increasing function of $\mathcal{S}$, positivity, and 
Bekenstein-Hawking limit.

To recap, we have proposed two new examples of entropy that may be valid 
for the description of certain physical systems, which we have not yet 
discussed.  Eventually, several additional proposals for even more general 
entropies can be conceived. However, we still lack a physical principle 
selecting an entropy as unique and universal, perhaps containing many 
parameters depending on various quantities.

\section{Black hole thermodynamics with 3-parameter generalized entropy}
\label{sec:3}

It is interesting to see what happens when the generalized 
entropy~(\ref{general}) is ascribed to the prototypical black hole, given 
by the Schwarzschild geometry \cite{Wald}
\begin{align}
\label{dS3BB}
ds^2 = - f(r) \, dt^2 + \frac{dr^2}{f(r)} + r^2 d\Omega^2_{(2)}\, , 
\quad\quad  
f(r) = 1 - \frac{2GM}{r} \,,
\end{align}
where  $M$ is the black hole mass and $d\Omega^2_{(2)}=d\vartheta^2 
+\sin^2 \vartheta \, d\varphi^2$ is the line element on  the unit 
two-sphere.  The black hole event horizon is located at the Schwarzschild 
radius 
\begin{align}
\label{horizonradius}
r_\mathrm{H}=2GM\, .
\end{align}
Studying quantum field theory on the spacetime with this horizon, 
Hawking discovered that the Schwarzschild black hole radiates with a 
blackbody spectrum at the temperature \cite{Hawking:1975vcx}  
\begin{align}
\label{dS6BB}
T_\mathrm{H} = \frac{1}{8\pi GM}\,.
\end{align}
In Ref.~\cite{Nojiri:2021czz} we attempted to identify the Tsallis 
entropy~(\ref{TS1}) or the R{\' e}nyi entropy~(\ref{RS1}) with the black 
hole entropy. As explained in general below, if we assume that the mass 
$M$ coincides with the thermodynamical energy, then the temperature 
obtained from the thermodynamical law is different from the Hawking 
temperature, a contradiction for observers detecting Hawking radiation. 
Alternatively, if the Hawking temperature $T_\mathrm{H}$ is identified 
with the physical black hole temperature, the obtained thermodynamical 
energy differs from the Schwarzschild mass $M$ even for the Tsallis 
entropy or the R{\' e}nyi entropy, which seems to imply a breakdown of 
energy conservation. Below, we follow the same procedure employed 
in Ref.~\cite{Nojiri:2021czz}.

If the mass $M$ coincides with the thermodynamical energy $E$ of the 
system due to energy conservation, as in 
\cite{Czinner:2015eyk,Tannukij:2020njz,Promsiri:2020jga,Samart:2020klx},  
in order for this system to be consistent with the thermodynamical 
equation $d\mathcal{S}_G=dE/T$ one needs to define the 
generalized temperature $T_\mathrm{G}$ as 
\begin{align}
\label{TR1}
\frac{1}{T_\mathrm{G}} \equiv \frac{d\mathcal{S}_\mathrm{G}}{dM} 
\end{align}
which is, in general, different from the Hawking temperature $T_\mathrm{H}$. 
For example, in the case of the entropy~(\ref{general6}), one 
has 
\begin{align}
\label{TR1B}
\frac{1}{T_\mathrm{G}} 
= \frac{\alpha}{\gamma} \left( 1 + \frac{\alpha}{\beta} \, \mathcal{S} 
\right)^{\beta-1} \frac{d\mathcal{S}}{dM} 
= \frac{\alpha}{\gamma}  \left( 1 + \frac{\alpha}{\beta} \, \mathcal{S} 
\right)^{\beta-1} \frac{1}{T_\mathrm{H}} \,,
\end{align}
where 
\begin{align}
\label{SMT}
\mathcal{S}=\frac{A}{4G}= 4 \pi G M^2 = \frac{1}{16\pi G{T_\mathrm{H}}^2} 
\,.
\end{align}
Because $\frac{\alpha}{\gamma} \left( 1 + 
\, \frac{\alpha}{\beta} \, \mathcal{S} 
\right)^{\beta-1} \neq 1$, it is necessarily $T_\mathrm{G}\neq 
T_\mathrm{H}$. Since the Hawking temperature~(\ref{dS6BB}) is the 
temperature perceived by observers detecting Hawking radiation, the 
generalized temperature $T_\mathrm{G}$ in (\ref{TR1B}) cannot be a  
physically meaningful temperature.

In Eq.~(\ref{TR1}), assuming that the thermodynamical energy $E$ is the 
black hole mass $M$ leads to an unphysical result. As an 
alternative, assume that the thermodynamical temperature $T$ coincides 
with the Hawking temperature $T_\mathrm{H}$ instead of assuming $E=M$. 
This assumption leads to 
\begin{align}
\label{Energy}
dE_\mathrm{G} = T_\mathrm{H} \, d\mathcal{S}_\mathrm{G}
= \frac{d\mathcal{S}_\mathrm{G}}{d\mathcal{S}} \,  
\frac{d\mathcal{S}}{\sqrt{16\pi G \mathcal{S}}} 
\end{align}
which, in the case of Eq.~(\ref{general6}), yields 
\begin{align}
\label{Energy2}
dE_\mathrm{G} = \frac{\alpha}{\gamma} \left( 1 + 
\frac{\alpha}{\beta} \, \mathcal{S} \right)^{\beta-1} \, 
\frac{d\mathcal{S}}{\sqrt{16\pi G \mathcal{S}}} 
= \frac{\alpha}{\gamma \sqrt{16\pi G}} \left[ \mathcal{S}^{-1/2} 
+ \frac{\alpha \left( \beta -1 \right)}{\beta} \, \mathcal{S}^{1/2} + 
\mathcal{O}\left( \mathcal{S}^{3/2} \right) \right] \, .
\end{align}
The integration of Eq.~(\ref{Energy2}) gives
\begin{align}
\label{GE2}
E_\mathrm{G} = \frac{\alpha}{\gamma \sqrt{16\pi G}} \left[ 2 
\mathcal{S}^{1/2} 
+ \frac{2\alpha \left( \beta -1 \right)}{3\beta} \, \mathcal{S}^{3/2} 
+ \mathcal{O}\left( \mathcal{S}^{5/2} \right) \right]  
= \frac{\alpha}{\gamma} \left[ M + \frac{4\pi G \alpha \left( \beta -1 
\right)}{3\beta} M^3 + \mathcal{O}\left( M^5 \right) \right] \,, 
\end{align}
where the integration constant is determined by the condition that 
$E_\mathrm{G}=0$ when $M=0$. Even when $\alpha=\gamma$, due to the 
correction $\frac{4\pi G \alpha \left( \beta -1 \right)}{3\beta} M^3$, the 
expression~(\ref{GE2}) of the thermodynamical energy $E_\mathrm{R}$ 
obtained differs from the black hole mass $M$, $E_\mathrm{G}\neq E$, which 
seems unphysical. In Einstein gravity, we always find $E=M$ for the 
Schwarzschild black hole. We may consider a process in which the 
Schwarzschild black hole forms from the collapse of a sufficiently large 
spherically symmetric shell of dust with mass $M$. In this process, the 
thermodynamical energy $E$ should initially be equal to the mass $M$ of 
the dust, $E=M$. In Einstein gravity, the Jebsen-Birkhoff theorem 
\cite{Wald} forces the spacetime outside the shell to be the Schwarzschild 
one~(\ref{dS3BB}), where $M$ in Eq.~(\ref{dS3BB}) is the shell mass.  
Inside this shell, spacetime is empty and flat, due again to the 
Jebsen-Birkhoff theorem \cite{Wald}. A black hole is formed by the 
collapse of the shell when the radius of the latter becomes smaller than 
its Schwarzschild radius~(\ref{horizonradius}).  If energy is conserved 
because the geometry outside the shell is not changed, the thermodynamical 
energy $E$ must be equal to $M$ after the formation of the black hole 
event horizon. In general, the Jebsen-Birkhoff theorem does not hold in 
theories of gravity extending general relativity 
\cite{Willbook,Will:2014kxa}, hence the geometry outside the shell is not 
always forced to be the Schwarzschild one and there might be emission of 
energy through the radiation of scalar modes. Therefore, $E\neq M$ might 
signal a theory of gravity beyond Einstein gravity.

One can verify that the six-parameter generalized 
entropy~(\ref{general1}), as well, seems inconsistent with the description 
of black hole thermodynamics.

\section{Holographic cosmology with generalized entropy}
\label{sec:4}

Dark energy models motivated by holography have been the subject of a 
considerable amount of literature ({\em e.g.},  
\cite{Li:2004rb,Li:2011sd,Wang:2016och, 
Nojiri:2005pu,Enqvist:2004xv,Zhang:2005yz,Guberina:2005fb,Elizalde:2005ju, 
Ito:2004qi,Gong:2004cb,Saridakis:2007cy,Gong:2009dc,BouhmadiLopez:2011xi,Malekjani:2012bw,Khurshudyan:2014axa,Khurshudyan:2016gmb,Landim:2015hqa, 
Gao:2007ep,Li:2008zq,Anagnostopoulos:2020ctz,Zhang:2005hs,Li:2009bn,Feng:2007wn,Lu:2009iv, 
Micheletti:2009jy,Huang:2004wt,Mukherjee:2017oom,Nojiri:2017opc,Sharif:2019seo,Jawad:2018juh}). 
The density of the holographic dark energy (HDE) is proportional 
to the square of 
the inverse holographic cutoff $L_\mathrm{IR}$,
\begin{equation}
\label{basic}
\rho_\mathrm{hol}=\frac{3C^2}{\kappa^2 {L_\mathrm{IR}}^2}\,,
\end{equation}
where $C$ is a free parameter. The holographic cutoff $L_\mathrm{IR}$ is 
usually assumed to 
be the same as the particle horizon $L_\mathrm{p}$ or the future horizon 
$L_\mathrm{f}$. No compelling argument has been proposed thus far for 
choosing this quantity, hence the most general cutoff was 
proposed in Ref.~\cite{Nojiri:2005pu}. 
In this proposal, the cutoff is assumed to depend upon 
$L_\mathrm{IR} = L_\mathrm{IR}(L_\mathrm{p}, \dot L_\mathrm{p}, \ddot 
L_\mathrm{p}, \cdots, L_\mathrm{f}, \dot L_\mathrm{f}, \cdots, a)$, which 
in turn leads to the generalized version of HDE known as ``generalized 
HDE'' \cite{Nojiri:2005pu,Nojiri:2021iko,Nojiri:2021jxf}.
In the spatially flat Friedmann-Lema\^{i}tre-Robertson-Walker 
(FLRW) universe described by the line element 
\begin{equation}
ds^2=-dt^2+a^2(t)\sum_{i=1}^{3} \left(dx^i\right)^2 
\label{metric}
\end{equation}
with scale factor $a(t)$ in comoving coordinates $\left( t,x,y,z \right)$, 
one might speculate that the generalized 
HDE  originates from one of several kinds of entropies associated with the 
cosmological horizon. In the FLRW spacetime~(\ref{metric}), the particle 
horizon  $L_\mathrm{p}$ and the future event horizon $L_\mathrm{f}$ are 
defined as
\begin{equation}
\label{H3}
L_\mathrm{p}\equiv a(t) \int_0^t\frac{dt'}{a(t')}\ ,\quad\quad
L_\mathrm{f}\equiv a(t) \int_t^\infty \frac{dt'}{a(t')}\,,
\end{equation}
respectively, when these integrals converge. 
Differentiating both sides of these definitions leads to the expressions 
of the Hubble function in terms of $L_\mathrm{p}$, $\dot{L}_\mathrm{p}$ or  
of $L_\mathrm{f}$, $\dot{L}_\mathrm{f}$ (where an overdot denotes 
differentiation with respect to the comoving time $t$)
\begin{equation}
\label{HLL}
H \left( L_\mathrm{p} , \dot{L}_\mathrm{p} \right) =
\frac{\dot{L}_\mathrm{p}}{L_\mathrm{p}} - \frac{1}{L_\mathrm{p}}\, , 
\quad \quad 
H(L_\mathrm{f} , \dot{L}_\mathrm{f}) = 
\frac{\dot{L}_\mathrm{f}}{L_\mathrm{f}} + \frac{1}{L_\mathrm{f}} \,,
\end{equation}
where the Hubble rate is $H\equiv \dot a/a $. 

As argued, {\em e.g.}, in Ref.~\cite{Padmanabhan:2009vy}, the standard 
Einstein-Friedmann equations can be derived from the Bekenstein-Hawking 
entropy~(\ref{SAdS10}). The physical radius of the cosmological horizon 
in spatially flat FLRW universes is 
\begin{equation}
\label{apphor}
r_\mathrm{H}=\frac{1}{H}\, ,
\end{equation}
which tells us that the entropy inside this  
horizon can be given by the Bekenstein-Hawking entropy~(\ref{SAdS10}) with 
the identification $A\equiv 4\pi {r_\mathrm{h}}^2=4\pi {r_\mathrm{H}}^2$. 
Because the incremental change of the energy $E$, or the increase of the 
heat $Q$, contained in this region is given by
\begin{equation}
\label{Tslls2}
dQ = - dE = -\frac{4\pi}{3} \, r_\mathrm{H}^3 \, \dot{\rho} \, dt = 
-\frac{4\pi}{3H^3} \, \dot{\rho} \,  dt  = \frac{4\pi}{H^2} \left( \rho + 
P \right) dt 
\end{equation}
(where we used the conservation law $ \dot \rho + 3 H \left( \rho + P 
\right)=0$), by using the Gibbons-Hawking temperature 
\cite{Cai:2005ra}
\begin{equation}
\label{Tslls6}
T = \frac{1}{2\pi r_\mathrm{H}} = \frac{H}{2\pi}
\end{equation}
and the first law of thermodynamics $TdS = dQ$, 
we obtain 
\begin{align}
\label{dotH}
\dot H = - 4\pi G \left( \rho + P \right) \, .
\end{align}
The integration of Eq.~(\ref{dotH}) leads to  the Friedmann equation 
\begin{equation}
\label{Tslls8}
H^2 = \frac{8\pi G}{3} \rho + \frac{\Lambda}{3} \,,
\end{equation}
where the integration constant corresponds to the cosmological 
constant $\Lambda$. 

It is possible to derive the black hole entropy from holography. As shown 
below, if we replace the Bekenstein-Hawking entropy~(\ref{SAdS10}) with 
another entropy and we apply the procedure illustrated between 
Eqs.~(\ref{Tslls2}) and~(\ref{Tslls8}), the Friedmann 
equation~(\ref{Tslls8}) is modified and extra contributions, which can be 
seen as holographic dark energy, arise from the non-standard entropy. For 
example, if we use the Tsallis entropy~(\ref{TS1}) instead of the 
Bekenstein-Hawking entropy~(\ref{SAdS10}), Eq.~(\ref{dotH}) is modified to  
\begin{align}
\delta\left(\frac{H}{H_1}\right)^{2(1-\delta)}\dot{H} = - 4\pi G 
\left( \rho + P \right)\,, \label{Tsallis-new1}
\end{align}
where ${H_1}^2 \equiv 4\pi/A_0 $. The integration of  
Eq.~(\ref{Tsallis-new1}) yields
\begin{align}
H^2=\frac{8\pi G}{3} \left( \rho + \rho_\mathrm{T} \right) + 
\frac{\Lambda}{3} \, , \quad 
\rho_\mathrm{T} = \frac{3}{8\pi G} 
\left[ H^2 - \frac{\delta}{2 - \delta}\, H_1^2\left( \frac{H}{H_1} 
\right)^{2(2 - \delta)} \right] \, .
\label{Tslls11BB}
\end{align}  
If we interpret $\rho_\mathrm{T}$ as the holographic dark energy due to  
 the holographic infrared cutoff $L_\mathrm{IR,T}$,  
$\rho_\mathrm{T}=\frac{3C^2}{\kappa^2 {L_\mathrm{IR,T}}^2}$, 
then the holographic infrared cutoff $L_\mathrm{IR,T}$ can be identified 
with 
\begin{align}
\label{Tbasic}
L_\mathrm{IR,T} =& \frac{1}{C\sqrt{ H^2 - \frac{\delta}{2 - \delta} 
H_1^2\left( \frac{H}{H_1} \right)^{2(2 - \delta)} }} \nonumber \\
=& \frac{1}{C\sqrt{ \left( \frac{\dot{L}_\mathrm{p}}{L_\mathrm{p}} - 
\frac{1}{L_\mathrm{p}} \right)^2 
 - \frac{\delta}{2 - \delta} \, H_1^2\left( 
\frac{\frac{\dot{L}_\mathrm{p}}{L_\mathrm{p}} - 
\frac{1}{L_\mathrm{p}}}{H_1} \right)^{2(2 - \delta) } }} \nonumber\\
 = & \frac{1}{C\sqrt{ \left( \frac{\dot{L}_\mathrm{f}}{L_\mathrm{f}} + 
\frac{1}{L_\mathrm{f}} \right)^2 
 - \frac{\delta}{2 - \delta} \, H_1^2\left( 
\frac{\frac{\dot{L}_\mathrm{f}}{L_\mathrm{f}} + 
\frac{1}{L_\mathrm{f}}}{H_1} \right)^{2(2 -\delta) } }} \, .
\end{align}
Equivalently, such a FLRW equation can always be rewritten in 
terms of a generalised cosmological dark fluid 
(see \cite{Bamba:2012cp} for a review). 
A similar procedure for the R{\'e}nyi entropy~(\ref{RS1}) gives 
\begin{align}
\label{Re2}
\rho_\mathrm{R} =& \, 
\frac{3 \pi \alpha}{8 G^2} \ln \left( 1 + \frac{G H^2}{\pi \alpha} 
\right) \, .
\end{align}

In the case of the Sharma-Mittal entropy~(\ref{SM}), if we simplify the 
situation by replacing the Tsallis entropy $\mathcal{S}_\mathrm{T}$ in 
Eq.~(\ref{SM}) with the Bekenstein-Hawking entropy $\mathcal{S}$ 
in~(\ref{SAdS10}) contained in it as a limit, we obtain
\begin{align}
\label{SM3}
\rho_\mathrm{SM} = \frac{3}{8\pi G}
\left[ H^2 - \frac{\pi}{G\left(2 - R/\delta \right)} 
\left(\frac{G H^2}{\pi} \right)^{2 - R/\delta } 
{}_2F_1 \left( 1 - \frac{R}{\delta}, 2 - \frac{R}{\delta}, 3 - 
\frac{R}{\delta}; 
 -\frac{G H^2}{\pi} \right) \right] \,,
\end{align}
where ${}_2F_1 \left(a,b, c; z \right) $ is the hypergeometric function. 
For the Barrow entropy~(\ref{barrow}), one obtains instead
\begin{align}
\label{rhoB}
\rho_\mathrm{B} = \frac{3}{8\pi G} 
\left[ H^2 - \left( \frac{1 + \Delta/2}{1 - \Delta/2} \right)   
\frac{16 \pi G}{A_\mathrm{Pl}^2} \left( \frac{H^2}{4\pi A_\mathrm{Pl}} 
\right)^{1 - \Delta/2 } \right] \, .
\end{align}

The three-parameter entropy~(\ref{general6}) gives 
\begin{align}
\label{3G}
\rho_\mathrm{G} = \frac{3}{8\pi G}
\left[ H^2 - \frac{\pi\alpha}{G\beta\gamma\left( 1 - \beta\right)} 
\left(\frac{G\beta H^2}{\pi \alpha} \right)^{2 - \beta} 
{}_2F_1 \left( 1 - \beta, 2 - \beta, 3 - \beta;  - \frac{G \beta 
H^2}{\pi\alpha} \right) \right] \, ,
\end{align}
which is expressed in terms of the particle horizon $L_\mathrm{p}$ or the 
future event horizon $L_\mathrm{f}$ by 
\begin{eqnarray}
\label{3G2}
\rho_\mathrm{G} &=& \frac{3}{8\pi G}
\left[ \left( \frac{\dot{L}_\mathrm{p}}{L_\mathrm{p}} - 
\frac{1}{L_\mathrm{p}} \right)^2 - \frac{\pi\alpha}{G\beta\gamma\left( 1 
- \beta\right)} 
\left(\frac{G\beta \left( \frac{\dot{L}_\mathrm{p}}{L_\mathrm{p}} - 
\frac{1}{L_\mathrm{p}} \right)^2}{\pi \alpha} \right)^{2 - \beta} 
\right.\nonumber\\
& \, &  \left.  \times 
{}_2F_1 \left( 1 - \beta, 2 - \beta, 3 - \beta; - \frac{G \beta \left( 
\frac{\dot{L}_\mathrm{p}}{L_\mathrm{p}} - \frac{1}{L_\mathrm{p}} 
\right)^2} {\pi\alpha} \right) \right] \nonumber \\
&\nonumber\\
& =& \frac{3}{8\pi G}
\left[ \left(  \frac{\dot{L}_\mathrm{f}}{L_\mathrm{f}} + 
\frac{1}{L_\mathrm{f}} \right)^2 - \frac{\pi\alpha}{G\beta\gamma\left( 1 
- \beta\right)} 
\left(\frac{G\beta \left( \frac{\dot{L}_\mathrm{f}}{L_\mathrm{f}} + 
\frac{1}{L_\mathrm{f}} \right)^2}{\pi \alpha} \right)^{2 - \beta} 
\right.\nonumber\\
&\, & \left. \times 
{}_2F_1 \left( 1 - \beta, 2 - \beta, 3 - \beta; - \frac{G \beta  \left( 
\frac{\dot{L}_\mathrm{f}}{L_\mathrm{f}} + \frac{1}{L_\mathrm{f}} \right)^2}
{\pi\alpha} \right) \right] \,,
\end{eqnarray}
where the hypergeometric series terminates and reduces to a polynomial  if 
$\beta$ is an integer $ m\geq 1$.  One can define the pressure of the 
holographic dark energy $P_\mathrm{G}$ 
by means  of the covariant conservation law
\begin{align}
\label{conservation}
\dot\rho_\mathrm{G} + 3 H \left( \rho_\mathrm{G} + P_\mathrm{G} 
\right) =0 \,;
\end{align}
the equation of state parameter $w_\mathrm{G}$ can then be written as 
\begin{align}
\label{eos}
w_\mathrm{G} \equiv &\, \frac{P_\mathrm{G}}{\rho_\mathrm{G}} 
= - 1 - \frac{\dot\rho_\mathrm{G}}{3 H \rho_\mathrm{G}} \nonumber \\
=& -1 - \frac{2}{3} \dot H \left[ H^2 - 
\frac{\pi\alpha}{G\beta\gamma\left( 1 - \beta\right)} 
\left(\frac{G\beta H^2}{\pi \alpha} \right)^{2 - \beta} 
{}_2F_1 \left( 1 - \beta, 2 - \beta, 3 - \beta; - \frac{G \beta 
H^2}{\pi\alpha} \right) \right]^{-1} \nonumber \\
& \times \left[ 1 - \frac{2 - \beta}{\gamma\left( 1 - \beta\right)} 
\left(\frac{G\beta H^2}{\pi \alpha} \right)^{1 - \beta} 
{}_2F_1 \left( 1 - \beta, 2 - \beta, 3 - \beta; - \frac{G \beta 
H^2}{\pi\alpha} \right) \right. \nonumber \\
& \qquad \left. + \frac{2- \beta}{\gamma\left( 3 - \beta\right)} 
\left(\frac{G\beta H^2}{\pi \alpha} \right)^{2 - \beta} 
{}_2F_1 \left( 2 - \beta, 3 - \beta, 4 - \beta; - \frac{G \beta 
H^2}{\pi\alpha} \right) \right] \,. 
\end{align}
When the matter contribution is negligible and the 
cosmological constant vanishes, the Friedmann equation reads  
\begin{align}
\label{HGF5}
H^2 = \frac{8\pi G}{3} \rho_\mathrm{G} 
\end{align}
and then Eq.~(\ref{3G}) gives 
\begin{align}
\label{HGF6}
{}_2F_1 \left( 1 - \beta, 2 - \beta, 3 - \beta; - \frac{G \beta 
H^2}{\pi\alpha} \right) =0 \, .
\end{align}
Therefore, the zeros $Z_i$ of the hypergeometric function ${}_2F_1 
\left( 1 - \beta, 2 - \beta, 3 - \beta; z \right)$ correspond to de Sitter 
universes with Hubble constant $H$ given by 
\begin{align}
\label{HGF7}
Z_i = - \frac{G \beta H^2}{\pi\alpha} \, .
\end{align}
Then, in spite of the absence of a true cosmological constant 
$\Lambda$, 
Eq.~(\ref{HGF7}) gives the effective cosmological constant 
\begin{align}
\label{HGF7B}
\Lambda_\mathrm{eff} = \frac{3\pi \alpha Z_i}{G \beta} \, .
\end{align}
Since $H$ is constant ($ \dot H = 0 $), if $H$ is given 
by Eq.~(\ref{HGF7}) the equation of state parameter $w_\mathrm{G}$ in 
(\ref{eos}) is almost 
$-1$, $w_\mathrm{G} \sim -1$. If $\Lambda_\mathrm{eff}$ in (\ref{HGF7B}) 
is large, this effective cosmological constant may describe inflation. 
On the other hand, if $\Lambda_\mathrm{eff}$ is sufficiently small, the 
effective cosmological constant may describe the accelerated expansion of 
the present universe. If the effective cosmological constant is slightly 
larger than the present dark energy, this effective constant could 
potentially solve the Hubble tension problem.

Let us first consider the case in which $Z_i$ (which we now write as $Z_1$ 
for $  i=1 $) is sufficiently small.  When $\frac{G \beta 
H^2}{\pi\alpha}$ is 
small, the hypergeometric function  ${}_2F_1 \left( 1 - \beta, 2 - \beta, 
3 - \beta; - \frac{G \beta H^2}{\pi\alpha} \right)$ is expanded as 
\begin{eqnarray}
{}_2F_1 \left( 1 - \beta, 2 - \beta, 3 - \beta; - \frac{G \beta 
H^2}{\pi\alpha} \right)  
&=& 1 - \frac{\left( 1 - \beta \right) \left( 2 - \beta \right)}{3- 
\beta} \frac{G \beta H^2}{\pi\alpha}  \nonumber\\
&\, & + \frac{\left( 1 - \beta \right) \left( 2 - \beta \right)^2}{4 - 
\beta}\left(\frac{G \beta H^2}{\pi\alpha} \right)^2
+\mathcal{O} \left( \left( \frac{G \beta H^2}{\pi\alpha} \right)^3 
\right)\, . \label{HGF}
\end{eqnarray}
Therefore, if we neglect the terms of order $\mathcal{O} 
\left( \left( \frac{G \beta H^2}{\pi\alpha} \right)^2 \right)$ 
in Eq.~(\ref{HGF}) when $H$ is small, Eqs.~(\ref{HGF}) and (\ref{HGF6}) 
give 
\begin{align}
\label{HGF7C}
Z_1 = - \frac{G \beta H^2}{\pi\alpha} \sim - \frac{\left(3- 
\beta\right)}{\left( 1 - \beta \right) \left( 2 - \beta \right)} \, ,
\end{align}
that is, 
\begin{align}
\label{HGF8}
H^2 \sim \frac{\left(3- \beta\right)\pi\alpha }{\left( 1 - \beta \right) 
\left( 2 - \beta \right)G \beta}  
\end{align}
which becomes small when $\beta \lesssim 3$ and the terms of 
order $\mathcal{O} \left( 
\left( \frac{G \beta H^2}{\pi\alpha} \right)^2 \right)$ in 
Eq.~(\ref{HGF}) can be dropped.  This conclusion hints at the idea that 
the  solution~(\ref{HGF8}) could explain dark energy in the present 
universe.  We may assume 
\begin{align}
\label{HGF8B}
3 - \beta \sim \mathcal{O}\left( 10^{-2n} \right)\, , \quad \alpha \sim 
\mathcal{O}\left( 10^{-2m} \right)\, , 
\end{align}
and then Eq.~(\ref{HGF8}) gives
\begin{align}
\label{HGF9}
H^2 \sim \left( 10^{-n-m + 28}\, \mathrm{eV} \right)^2 \, ;
\end{align}
therefore, if $n+m= 61$, it is $H\sim 10^{-33}\, \mathrm{eV}$, which 
reproduces the present energy scale  of the dark energy. If  
another zero $Z_2$ exists with absolute value slightly smaller than $Z_1$, 
the  effective cosmological constant can potentially solve the Hubble 
tension problem, {\em i.e.}, the recent observational tension between the 
value of the Hubble constant inferred from small redshifts (as in the 
observations of Type Ia supernova calibrated by Cepheids  
\cite{Riess:2020fzl}) and that from large redshifts inferred from the 
cosmic microwave background (CMB) \cite{Planck:2018vyg}. This problem 
might be solved, or at least alleviated, if there is effectively dark 
energy just after the CMB was emitted. 
Our model admitting two zeros $Z_{1,2}$ with  $\left| Z_2 \right|$ 
slightly larger than $\left| Z_1 \right|$ might play the role of the 
effective dark energy just after the CMB.

In general, the hypergeometric function can have several or even 
an infinite number of 
zeros. If there are a root of  order unity or a large and negative root  
$Z_i$ of the equation $ {}_2F_1 \left( 1 - \beta, 2 - \beta, 3 - \beta; 
Z_i \right) =0$, then Eq.~(\ref{HGF7}) can give the large Hubble rate $H$ 
corresponding to the inflationary epoch. The 
Hubble rate $H$ and the effective cosmological constant 
$\Lambda_\mathrm{eff}$ are  
given by Eqs.~(\ref{HGF7}) and~(\ref{HGF7B}), respectively. If, for the  
sake of illustration, we retain the first three terms in Eq.~(\ref{HGF}), 
the latter assumes the form 
\begin{align}
\label{HGF10}
1 - \frac{\left( 1 - \beta \right) \left( 2 - \beta \right)}{3- 
\beta} \frac{G \beta H^2}{\pi\alpha}  + \frac{\left( 1 - \beta \right) 
\left( 2 - \beta \right)^2}{4 - \beta} \left(\frac{G \beta H^2}{\pi\alpha} 
\right)^2 =0 
\end{align}
with solutions 
\begin{align}
\label{HGF11}
\frac{G \beta H^2}{\pi\alpha} = Z_\pm 
\equiv& -\frac{ \frac{\left( 1 - \beta \right) \left( 2 - \beta 
\right)}{3- \beta} 
\pm \sqrt{\frac{\left( 1 - \beta \right)^2 \left( 2 - \beta 
\right)^2}{\left(3- \beta\right)^2} - 4 \, \frac{\left( 1 - \beta \right) 
\left( 2 - \beta \right)^2}{4 - \beta} } }
{\frac{2\left( 1 - \beta \right) \left( 2 - \beta \right)^2}{4 - \beta}} 
\nonumber \\
= & - \frac{4 - \beta}{2\left( 2 - \beta \right)\left( 3- \beta \right)}
\left( 1 \pm \sqrt{1 - \frac{4 \left( 3 - \beta \right)^2}{\left(4 - 
\beta\right)\left( 1 - \beta \right)} } \, \right) \, .
\end{align}
As in Eqs.~(\ref{HGF8B}) and (\ref{HGF9}) we assume $\beta \lesssim 3$, 
obtaining
\begin{align}
Z_+ = - \frac{4 - \beta}{2\left( 2 - \beta \right)\left( 3- \beta 
\right)} \, , \quad \quad
Z_- = - \frac{3 - \beta}{2\left( 1 - \beta \right)\left( 2- \beta 
\right)} 
\end{align}
(here $Z_-$ corresponds to $Z_1$ in Eq.~(\ref{HGF7C})). 
Therefore, if one writes $\alpha$ and $\beta$ as in Eq.~(\ref{HGF8B}) and 
chooses $n+m= 61$ as done below Eq.~(\ref{HGF9}),  
one finds again a Hubble constant $H$ that reproduces the present value of 
the dark energy scale.  If, instead, $\frac{G \beta H^2}{\pi\alpha} = 
Z_+$, one finds 
\begin{align}
\label{HGF12}
H^2 \sim \left( 10^{n-m + 28}\, \mathrm{eV} \right)^2 
\end{align}
and the choice $n+m= 61$ gives 
\begin{align}
\label{HGF13}
H^2 \sim \left( 10^{-2m + 89}\, \mathrm{eV} \right)^2 \, .
\end{align}
Assuming GUT scale $\left( \sim 10^{16}\, \mathrm{GeV} = 10^{25}\, 
\mathrm{eV} \right)$ inflation $H\sim 10^{2\times 25 - 28}\, \mathrm{eV} = 
10^{22}\, \mathrm{eV}$, we obtain $m\sim 33$ or $34$. Therefore $Z_+$ may 
produce the inflationary epoch of the early universe.

Similarly, one can consider generalized HDE coming from our six-parameter 
entropy~(\ref{general1}): then, there are many more possibilities to 
realize realistic cosmic histories by choosing appropriately the 
corresponding parameters.

\section{Conclusions}
\label{sec:5}

New and old definitions of entropy abound in the literature, mostly 
arising from non-extensive statistical mechanics and thermodynamics or 
from quantum gravity, and varying according to the physical theory or the 
physical systems considered. Not surprisingly, a recurrent feature is the 
presence of long-range forces, due to which the partition function 
diverges and the Boltzman-Gibbs entropy fails. Here we have discussed 
black holes and the holographic universe as physical systems and we have 
proposed two generalized entropies that satisfy certain basic 
requirements: each of these generalized entropies $\mathcal{S}_G$ must vanish 
when $\mathcal{S}$ vanishes, must be positive-definite, and must reduce to 
the Bekenstein-Hawking entropy~(\ref{SAdS10}) in some limit. It is clear 
that some requirements must be imposed on generalized entropies to 
restrict the range of possible proposals, and the three conditions we 
impose seem minimal requirements. It is quite possible that the spectrum 
of generalized entropies that they allow is still too wide and that is 
should be 
restricted further. In the meantime, we adopt two proposals for 
generalized entropies (given by Eqs.~(\ref{general1}) 
and~(\ref{general6})), which are very general yet directly linked to the 
physics explored in many recent works. (One could add extra terms 
to~(\ref{general1}), but that would mean adding extra parameters and these 
terms would have to be set to zero anyway to reproduce, {\em e.g.}, the 
Kaniadakis entropy~(\ref{kani}).)

Both generalized entropies~(\ref{general1}) and~(\ref{general6}) satisfy 
the three criteria above and reproduce a variety of entropy notions 
introduced in the literature, including the R\'enyi \cite{renyi}, Tsallis 
\cite{tsallis}, Sharma-Mittal \cite{SayahianJahromi:2018irq}, Kaniadakis 
\cite{Kaniadakis:2005zk,Drepanou:2021jiv}, Barrow \cite{Barrow:2020tzx}, 
and Majhi's Loop Quantum Gravity \cite{Majhi:2017zao, Czinner:2015eyk, 
Mejrhit:2019oyi, Liu:2021dvj} entropy proposals. Although we have 
restricted our attention to gravitational systems such as black holes and 
cosmology, our prescriptions are potentially much more general and could 
be applied to many other systems of interest in statistical mechanics, 
information theory, and other areas of physics in which long-range 
interactions are present, but not only. These applications will be the 
subject of future research.

\section*{Acknowledgments}

This work is supported, in part, by MINECO (Spain), project 
PID2019-104397GB-I00 (S.~D.~O), by JSPS Grant-in-Aid for Scientific 
Research (C) No. 18K03615 (S.~N.), and by the Natural Sciences \& 
Engineering Research Council of Canada, Grant No. 2016-03803 (V.~F.).


\end{document}